\documentclass[twocolumn,showpacs,preprintnumbers,amsmath,amssymb,pra,superscriptaddress]{revtex4}

\usepackage{graphicx}
\usepackage{epsfig}
\usepackage{dcolumn}
\usepackage{textcomp}
\usepackage{bm}

\def\rm#1{\mathrm{#1}}
\def\bf#1{\mathbf{#1}}

\begin{document}


\title{Identifying an Experimental Two-State Hamiltonian to Arbitrary Accuracy} 

\author{Jared H. Cole}
 \email{j.cole@physics.unimelb.edu.au}
\affiliation{%
Centre for Quantum Computer Technology, School of Physics, The University of Melbourne, Melbourne, Victoria 3010, Australia.
}%
\author{Sonia G. Schirmer}
\affiliation{%
Department of Applied Mathematics and Theoretical Physics, University of Cambridge, Wilberforce Road, Cambridge, CB3 0WA, UK
}%
\author{Andrew D. Greentree}
\author{Cameron J. Wellard}
\affiliation{%
Centre for Quantum Computer Technology, School of Physics, The University of Melbourne, Melbourne, Victoria 3010, Australia.
}%
\author{Daniel K. L. Oi}
\affiliation{%
Department of Applied Mathematics and Theoretical Physics, University of Cambridge, Wilberforce Road, Cambridge, CB3 0WA, UK
}%
\author{Lloyd C. L. Hollenberg}
\affiliation{%
Centre for Quantum Computer Technology, School of Physics, The University of Melbourne, Melbourne, Victoria 3010, Australia.
}%

\date{\today}

\begin{abstract}
Precision control of a quantum system requires accurate determination of the effective system Hamiltonian.  We develop a method for estimating the Hamiltonian parameters for some unknown two-state system and providing uncertainty bounds on these parameters.  This method requires only one measurement basis and the ability to initialise the system in some arbitrary state which is not an eigenstate of the Hamiltonian in question.  The scaling of the uncertainty is studied for large numbers of measurements and found to be proportional to one on the square-root of the number of measurements.
\end{abstract}

\maketitle

\section{Introduction}
High precision control of quantum systems inevitably requires high precision characterisation of the system dynamics.  Quantum computers are an example of a device requiring especially high precision characterisation, but this precision is also required for detailed studies of interactions in quantum systems.  For qubits, this characterisation is usually performed using state and process tomography, where the full density matrix is measured for a range of different input states~\cite{Chuang:97,Poyatos:97,James:01,Ziman:04}.  An alternative approach is to directly characterise the Hamiltonian, which then gives the evolution of the system for any initial state.  This approach is especially useful when the system approaches a \textit{closed} system and therefore its dynamics can be treated as purely Hamiltonian.  While this will not be the case in general, it is an essential requirement for constructing a qubit for quantum computing applications and is approximately true for many other systems of interest.

Tomographic methods typically require measurement in several different bases or require the ability to perform rotations around particular axes before the system has been completely characterised.  In contrast, a general procedure developed recently for identifying an arbitrary two-state Hamiltonian\cite{Schirmer:04} requires measurement in only one basis and initialisation in a single known state.  The requirement for only one measurement basis is especially attractive for systems with limited measurement devices, for example many solid-state qubits.  We build on this result by deriving a systematic method to calculate the Hamiltonian parameters to any required accuracy from a time series of measurement data.  

In previous work~\cite{Schirmer:04} the Hamiltonian is assumed to be a linear combination of some free evolution Hamiltonian and various control fields, where characterisation requires finding both the base Hamiltonian and the dependence on the control field.  In this work, we take a more pragmatic approach to the problem of characterisation of a two-state system.  We provide a method to answer the question `What measurements must be taken to determine the form of a two-state Hamiltonian to a given precision?'  Assuming that the system evolves under some Hamiltonian, which corresponds to a certain `position on a dial' in the laboratory,  the parameters for this particular Hamiltonian can be determined to some arbitrary precision.  If the two-state system is to be used as a qubit for quantum information processing (QIP) applications, the process can then be repeated for some other linearly independent Hamiltonian, giving two `axes' that are sufficient to construct any arbitrary single-qubit rotation~\cite{Nielsen:00}.   
In general, the response of the system to various `dial settings' would be required to construct efficient single-qubit gates.  To do this, the Hamiltonian parameters and their uncertainty would need to be determined for a number of points and the response determined.  In the case of linear response, this becomes completely equivalent to the process discussed in reference~\cite{Schirmer:04} but more generally will require fitting to an appropriate functional form.

The basic outline of this method and the relevant equations are given in sections~\ref{sec:CharacProc}, \ref{sec:EstHamilParams} and \ref{sec:find_phi}.  The uncertainty in these estimates is then analysed and a series of uncertainty relations are given in sections~\ref{sec:theta_err},\ref{sec:phi_err} and \ref{sec:H_err} which allows the Hamiltonian to be estimated with error bounds on all its parameters.  Section \ref{sec:DFTbits} covers some technical details on the use of the discrete Fourier transform (DFT) to analyse the time series data and how its accuracy can be controlled for this particular application.  In section~\ref{sec:ExampleSims} we numerically simulate this method for some example Hamiltonians and compare the statistical spread of results with the estimated uncertainty, finding very good agreement.  We also investigate how the accuracy of the Hamiltonian parameters scales as a function of the number of measurements.  Finally, in section~\ref{sec:compositepulses} we discuss the effect of this scaling on the characterisation and operation of single qubit gates for QIP applications.

\section{Characterising a Two-State Hamiltonian}\label{sec:CharacProc}
\begin{figure} [tb!]
\centering{\includegraphics[width=7.5cm]{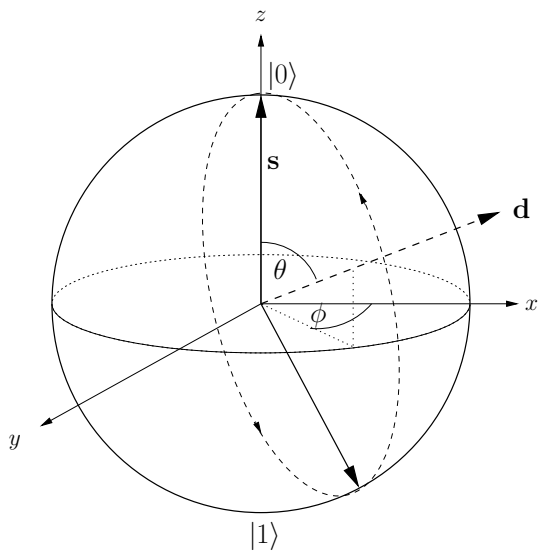}}
\caption{Bloch sphere representation of the state of a qubit and its trajectory given an arbitrary Hamiltonian $\bf{d}$.  If the system is not in an eigenstate of the Hamiltonian, the state given by the unit vector $\bf{s}$ precesses around an axis defined by $\bf{d}$.  The components of $\bf{d}$ are given by the Hamiltonian using Eq.~(\ref{eq:Handd}) where the $|\bf{d}|$ gives the angular precession frequency around the vector $(d_x,d_y,d_z)^T$.\label{fig:BlochSphere}}
\end{figure}

Some insight into the time evolution of an arbitrary superposition state can be gained by considering the Bloch sphere picture for a two-state system.  The Hamiltonian of an arbitrary two-level system can be written in terms of the Pauli matrices,
\begin{equation}\label{eq:Handd}
H=\frac{\bf{d}.\vec{\sigma}}{2}=\frac{|\bf{d}|}{2}(d_0I+d_x\sigma_x+d_y\sigma_y+d_z\sigma_z),
\end{equation}
where $d_x$,$d_y$ and $d_z$ are real constants and $d_0$ results in an unobservable global phase factor which can be ignored.  If the state of the system is mapped to the Bloch sphere, its position in the sphere is the Bloch vector ($\mathbf{s}$) where $|\bf{s}|\le1$, with a pure state having $|\bf{s}|=1$.  The evolution of the Bloch vector due to some Hamiltonian ($H$) will be to precess around a unit vector $(d_x,d_y,d_z)^T$ with angular rotation frequency given by $|\bf{d}|$.  If the system is in an eigenstate of the Hamiltonian, the Bloch vector is parallel to the axis of rotation and therefore does not precess, as expected.  This process is illustrated in Fig.~\ref{fig:BlochSphere}.  

\begin{figure} [tb!]
\centering{\includegraphics[width=8cm]{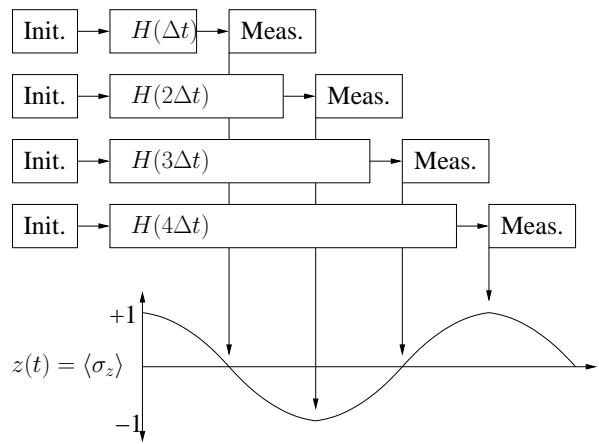}} 
\caption{To map $z(t)$ the system must be repeatedly initialised, allowed to evolve under the Hamiltonian to be measured $H(t)$ and then measured.  To map the time evolution of the system, the Hamiltonian step is applied for progressively longer time intervals ($i\Delta t$ for $i=1,2,\ldots,n$) where $\Delta t$ is the minimum controllable time interval and $t_{\rm{ob}}=n\Delta t$ is the maximum time over which the system is observed.\label{fig:ProcessDiag}}
\end{figure}

If the system can be repeatedly initialised in a known state and then measured in some basis at progressively longer time periods, the trajectory of the Bloch vector can be mapped.  Assuming this is an idealised projective measurement, the sinusoidal variation of the projection onto the measurement axis depends on both the magnitude and direction of the vector $\bf{d}$ and therefore on the parameters in the Hamiltonian.  A schematic of this process is shown in Fig.~\ref{fig:ProcessDiag} where the minimum controllable time interval is given by $\Delta t$ and the longest time the system is allowed to evolve is $t_{\rm{ob}}$ giving the total number of time points $N_s=t_{\rm{ob}}/\Delta t$.  This process is then repeated $N_e$ times to build up an ensemble average for each time point, giving a total of $N_T=N_e t_{\rm{ob}}/\Delta t=N_e N_s$ measurements.  The true evolution $z(t)$ is then approximated by the measured function $z_m(t)$. 

For simplicity, we will use polar coordinates to describe both the position of the Bloch vector and the Hamiltonian vector, as illustrated in Fig.~\ref{fig:BlochSphere}.  In these coordinates, the Hamiltonian vector is given by $\bf{d}=|\bf{d}|(d_x,d_y,d_z)^T=|\bf{d}|[\sin(\theta)\cos(\phi),\sin(\theta)\sin(\phi),\cos(\theta)]^T$.  As the complex phase ($\phi$) is unobservable in a single two-state system, we can set $\phi=0$ and therefore align the Hamiltonian with the x-axis.  If the reference axes are defined based on experimental grounds, this can be corrected with a trivial rotation.

If the system is initialised in the state $|\psi(0)\rangle=|0\rangle$ (which corresponds to $\theta=\phi=0$ or $\mathbf{s}_0=(0,0,1)^T$), the evolution of the z-component of the state vector is
\begin{equation}
z(t)=\cos(\omega t)\sin^2(\theta)+\cos^2(\theta),
\end{equation}
where $\omega=|\mathbf{d}|$ (in units such that $\hbar=1$), see reference~\cite{Schirmer:04} or appendix~\ref{app:sigz_derv} for an alternative derivation.  Determining the parameters $\omega$ and $\cos^2(\theta)$ gives the values of $|\mathbf{d}|$, $d_x$ and $d_z$.  Throughout this discussion, we assume that the Hamiltonian is constant in time and that the initial state of the system is not an eigenstate of the Hamiltonian, i.e.\ $\theta\neq0$, otherwise the system will not precess.  The process of characterising the Hamiltonian thus involves measuring $z_m(t)$ and analysing it to determine the appropriate parameters.

\section{Estimating the Hamiltonian parameters from Fourier components}\label{sec:EstHamilParams}
Once $z_m(t)$ is determined, the data can be fitted in the time domain to determine the Hamiltonian parameters\cite{Schirmer:04}.  While this is sufficient for approximate estimates or data containing only a few oscillation periods, a more elegant method is to take the Discrete Fourier Transform (DFT) of the data $z_m(t)$ and calculate the parameters from the Fourier coefficients.  This method provides both the Hamiltonian parameters and an estimate of the uncertainty in these values.  In order for this method to be effective, the Hamiltonian must be constant in time, or more precisely, the fields controlling the Hamiltonian must be stable to higher precision than that required for characterisation~\footnote{If the Hamiltonian has some random fluctuations, this will contribute to the noise level in the Fourier spectrum.  The magnitude of this `jitter' can be found by performing this analysis with progressively higher numbers of time-points.  If the uncertainty in the parameters is found to asymptote, this provides an estimate for the stability of the Hamiltonian.}.

As $z(t)$ is a pure sinusoid, if  $z_m(t)$ consists of an integer number of periods of oscillation, its Fourier transform ($F(\nu)=\rm{DFT}[z_m(t)]$) will  take on a simple form consisting of $\delta$-functions at $\nu=0$ and $\nu=\pm\nu_p$ where $\nu_p$ refers to the position of the peaks.  Using the definition of the inverse discrete Fourier transform $\rm{DFT}^{-1}$, $z_m(t)$ can be rewritten in terms of the discrete Fourier components for the zero [$F(0)$] and peak-frequencies [$F(\nu_p)$],
\begin{eqnarray*}
\rm{DFT}^{-1}[\rm{DFT}[z_m(t)]] 	& = 	& \sum_{\nu=-N_s/2}^{N_s/2}{F(\nu) e^{i2\pi(\nu/N_s)t}} \\
		& =	& F(0)+F(\nu_p)e^{i2\pi\nu_pt/N_s} \\ 
		&	& +F(-\nu_p)e^{-i2\pi\nu_pt/N_s} \\
		& =	& F(0)+2F(\nu_p)\cos(2\pi\nu_pt/N_s) \\
		& \simeq	& z(t).
\end{eqnarray*}
In this way, the angle $\theta$ and the angular precession frequency $\omega$ can be determined directly from the Fourier spectrum without the need for fitting the data in the time domain.

The effect of a measurement error probability can also be included by assuming some probability $\eta \in [0,1]$ of obtaining the incorrect value from a single measurement.  This corresponds to a bit-flip error ($\sigma_x$) occurring the instant before measurement with some probability $\eta$. Assuming the Bloch vector always starts at $|\psi(0)\rangle=|0\rangle$, $z(t)$ should reach a maximum of one after each period.  The measurement error will reduce this maximum, independent of the angle $\theta_r$ and can therefore be determined directly from the DFT.  If we model the effect of this measurement error as $z_m(t)=(1-2\eta) z(t)$, then the following equations can be derived,
\begin{equation}
\eta=\frac{1-F(0)}{2}-F(\nu_p),\label{eq:eta}
\end{equation}
\begin{equation}
\cos(\theta)=\sqrt{\frac{F(0)}{1-2\eta}},\label{eq:thetar}
\end{equation}
\begin{equation}
\omega=2\pi\nu_p/N_s.\label{eq:omega}
\end{equation}

\begin{figure} [tb!]
\centering{\includegraphics[width=8cm]{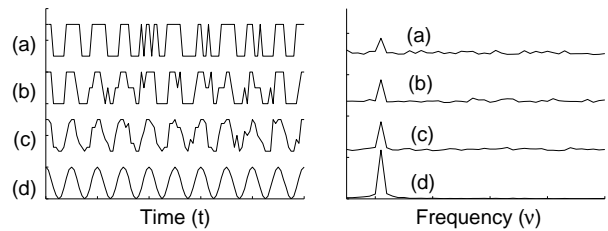}} \caption
{The left hand plot shows an example of a sampled time signal $z(t)=[\cos(2\pi t)+1]/2$ with $N_e=1$ (a), $2$ (b), $8$ (c) and $500$ (d) measurements at each time point, where each measurement is a projection onto the (1,-1) axis.  The corresponding DFT for each signal is shown on the right for $\nu\ge0$, illustrating the signal-to-noise improvement as more measurements are taken at each time point.\label{fig:dftexample}}
\end{figure}

As we can only perform projective measurements onto one axis, many measurements are required to accurately determine $z_m(t)$, so $N_T$ will typically be quite large.  Once the time resolution and observation time are chosen, the measurements for each time point can be repeated until a sufficiently resolved peak is seen in the DFT spectrum.  An example of this process is shown in Fig.~\ref{fig:dftexample} for progressively larger numbers of measurements at each time point.  In this way, the number of measurements need not be chosen at the start but the experiment is repeated until a sufficient signal-to-noise ratio is obtained.

\section{Determining the Precession Frequency to Arbitrary Accuracy}\label{sec:DFTbits}
Performing a discrete Fourier transform (DFT) on the measurement results immediately places some constraints on the selection of the measurement parameters.  In order to satisfy the Nyquist sampling criteria, at least two sample points for every period of oscillation are required to avoid aliasing.  This means that some estimate for the oscillation period $T_{\rm{predict}}$ must be known in order to guarantee that $\Delta t<T_{\rm{predict}}/2$, though in practice the period of oscillation will usually be known approximately on theoretical or experimental grounds.

Conventional DFT theory states that the frequency resolution ($\Delta \nu$) of a DFT signal is the inverse of half the total time of the signal, $\Delta \nu=2/t_{\rm{ob}}$\cite{Bracewell:00}.  This means that to resolve the frequency signal we need to observe at least two complete oscillation periods, though typically many more periods will need to be observed to obtain a clearly defined peak in the frequency spectra.  For an arbitrary signal the frequency resolution of the spectra also limits the precision with which one can determine the frequency ($\nu \pm \Delta \nu$).  The more periods observed the more accurately the determined frequency of oscillation.  Ultimately this will be restricted by the decoherence time of the system as decoherence reduces the amplitude of the oscillations for long observation times.

To use Eqs.~(\ref{eq:eta})-(\ref{eq:omega}), we require that the observation time $t_{\rm{ob}}$ is an integer number of periods.  To ensure this, we need to know the precession frequency to the same precision as the time control ($\Delta\nu/\nu_p \approx \Delta t / t_{\rm{ob}}$).  Conversely, if we can guarantee that we have an integer number of periods, this will yield the corresponding frequency.

\begin{figure} [tb!]
\centering{\includegraphics[width=8cm]{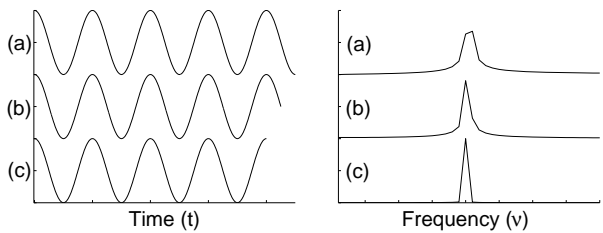}} \caption
{The left hand plot shows time signal which are truncated at various time points to produce a net phase difference of $\Delta\varphi=\pi$ (a), $\Delta\varphi=\pi/2$ (b) and $\Delta\varphi=0$ (c) between the start and end of the signal.  The corresponding DFT for each signal is shown on the right, where the peak approaches a $\delta$-function only for $\Delta\varphi \approx 0$.\label{fig:dftphase}}
\end{figure}

The DFT of a pure sinusoid has some special properties in that it only approaches a $\delta$-function when the time signal consists of an integer number of periods (there is no phase difference between the start and end of the signal)\cite{Bracewell:00}.  If there is some phase difference then the DFT has `leakage' into the other channels, resulting in a overall spread of the signal throughout the spectrum.  This effect is demonstrated in Fig.~\ref{fig:dftphase} for example sinusoids having various values for the phase difference ($\Delta\varphi=\varphi(0)-\varphi(t_{\rm{ob}})$) between the start and end points in the time signal.

Using this information, we can locate the `minimum-phase-point' (MPP) where the difference in phase between the start and end of the signal is minimised.  This amounts to selecting only an integer number of periods of the signal.  As the period of the signal is not known beforehand, the easiest method is to record the data and then reprocess it later to ignore some of the data points.  While this results in throwing away some information, the lost data consists of at most one period. 

\begin{figure} [tb!]
\centering{\includegraphics[width=5cm]{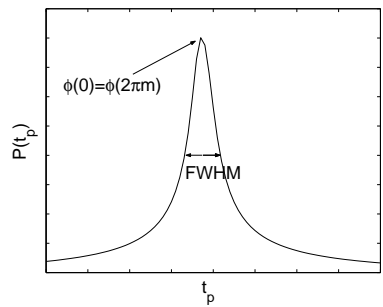}} \caption
{The test function $P(t_p)$ used to locate the point at which there is zero phase difference between the first and last sample point.  The amount of the time signal to use in the DFT is given by $t_p$ and the uncertainty is given by the FWHM of $P(t_p)$.\label{fig:Ptau}}
\end{figure}

An effective way of locating the MPP is to compare the magnitude of the channel comprising the central frequency peak $F(\nu_p)$ and its adjoining channels $F(\nu_p-1)$ and $F(\nu_p+1)$.  When the leakage is minimised, the ratio of the central channel to its neighbours should be a maximum.  An example test function which was found to perform well with varying levels of noise is
\begin{equation}
P(t_p)=\frac{2F(\nu_p)-F(\nu_p-1)-F(\nu_p+1)}{F(\nu_p-1)+F(\nu_p+1)},
\end{equation}
where once again $F(\nu)$ is the normalised DFT of the original signal from $z_m(0)$ to $z_m(t_p)$ where $t_{\rm{ob}}-T_{\rm{predict}}\le t_p \le t_{\rm{ob}}$.  An example plot of $P(t_p)$ is shown in Fig.~\ref{fig:Ptau}.  A clear peak is observed at the point where the phase of the sinusoid ($\varphi$) is an integer multiple of $2\pi$, i.e.\  $\varphi(0)=\varphi(t_p=2\pi m)$ for some integer $m$.  Once the MPP has been determined, the frequency is given by $\omega=2\pi n/t_p$ where $n$ is the peak channel number and $t_p$ is the MPP.
The advantage of this method is that the MPP can usually be determined to an accuracy of close to $\Delta t/t_{\rm{ob}}$ and the full-width-half-maximum (FWHM) of the function $P(t_p)$ gives an estimate for the uncertainty of the resulting frequency.

\section{Estimating the Uncertainty in the measured quantities}\label{sec:theta_err}
For most practical applications, if we wish to estimate the parameters of a two-state system, we also need to know the uncertainty in those estimates.  For the rest of the discussion we will use the following notation, $\hat{x}$ is the estimate obtained for some true value $x$ and $\delta x$ refers to the predicted standard deviation of this estimate.  In the ideal situation $x-3\delta x\le \hat{x} \le x+3\delta x$, with $99.7\%$ confidence.
 
As we are determining the parameters of interest from the components of the Fourier spectrum, we have a straightforward way of calculating the uncertainty from the spectral noise.  We define the noise spectrum $n(\nu)$ to be the parts of the Fourier spectrum which do not include $F(\pm\nu_p)$ and $F(0)$.  This is a good approximation when $t_{\rm{ob}}$ constitutes an integer number of periods and therefore $F(\pm\nu_p)$ and $F(0)$ approach $\delta$-functions. 

In general the noise due to the discrete measurement of the system will be a limiting factor in the analysis, though other factors like noise in the control Hamiltonian will also contribute.  The uncertainty in the frequency will be primarily controlled by the precision in the time control of the measurements.  Ideally the uncertainty in the angular frequency measurement should be of the same order as the time resolution in the measured signal ($\delta \omega/\omega \approx \Delta t/t_{\rm{ob}}$).  In practice a more accurate estimate for the uncertainty can be obtained from the FWHM of $P(t_p)$, as discussed earlier.  The uncertainty in the angle $\delta \theta$ and the measurement error $\delta \eta$ will be primarily limited by the noise level in the Fourier spectrum.  Typically, the fractional uncertainty in $\omega$ will be an order of magnitude smaller than for $\theta$ or $\eta$ as finding $\omega$ only requires finding the peak location where as the other parameters depend on the peak height which is directly affected by the spectral noise.

The uncertainty in the Fourier peaks is given by the standard deviation ($\rm{SD}$) of the noise spectrum.  For simplicity we will define $\delta F=\rm{SD}[n(\nu)]$ and $\delta\omega=2\pi/\rm{FWHM}[P(t_p)]$ so that the resulting uncertainty approximates the predicted standard deviation of the parameter estimate.  Once we have the uncertainty in the frequency $\delta \omega$ and the Fourier spectrum $\delta F$, using conventional uncertainty analysis~\cite{Kirkup:94} we can derive the expressions for the uncertainty in the calculated values.  Throughout this discussion we will use the standard error propagation method~\cite{Ku:69} where the variance of some function $w=f(x,y)$ is given in terms of the variances $\rm{var}(\hat{x})$ and $\rm{var}(\hat{y})$ and the covariance $\rm{cov}(x,y)$ between $x$ and $y$~\footnote{In this situation there is a correlation between the error in $F(0)$ and that in $F(\nu_p)$ as this error comes from the shared white-noise floor of the Fourier spectrum.  This means the covariance is not zero and is approximately equal to the variance of the noise signal itself.}.  In its simplest form, the variance of a function can be calculated using
\begin{eqnarray}
\rm{var}(\hat{w}) & = & \left[\frac{\partial F}{\partial X}\right]^2\rm{var}(\hat{x})+\left[\frac{\partial F}{\partial Y}\right]^2\rm{var}(\hat{y}) \nonumber \\ 
 & & +2\left[\frac{\partial F}{\partial X}\right]\left[\frac{\partial F}{\partial Y}\right]\rm{cov}(\hat{x},\hat{y}),
\end{eqnarray}
for small variances in the measured parameters. 

Using this approach, the uncertainty in each of the calculated quantities in Eq.~(\ref{eq:eta}) and (\ref{eq:thetar}) can be estimated using the following equations,
\begin{equation}\label{eq:m_err}
\delta \eta=\frac{3}{2}\delta F,
\end{equation}
\begin{eqnarray}
\delta A^2 & = & \frac{F(0)}{1-2\eta}\left[\left(\frac{\delta F}{2 F(0)}\right)^2+\left(\frac{\delta \eta}{1-2\eta}\right)^2\right] \nonumber \\
 & & +\left|\frac{1-2\eta-F(0)}{(1-2\eta)^3}\right|\delta F^2
\end{eqnarray}
and
\begin{equation}\label{eq:theta_err}
\delta \theta=\left|(1-A^2)^{-1/2}\right|\delta A,
\end{equation}
where $A=\cos(\theta)$.

This process results in an estimate and its associated uncertainty for the angular frequency $\omega$, rotation axis $\theta$ and the measurement error $\eta$.  A simplistic error analysis is given here to illustrate the ideas.  The use of more sophisticated techniques such as maximum likelihood estimation should provide tighter bounds on the estimated parameters for a given set of data\cite{James:01,DAriano:00,Fiurasek:01}. 

\section{Determining the phase-angle between two Hamiltonians}\label{sec:find_phi}
The process discussed so far is sufficient to characterise a single two-state Hamiltonian, as $d_y$ can be arbitrarily set to zero.  To provide a completely controllable two-state system, such as is needed for QIP, a second control Hamiltonian is required to implement all possible single-qubit rotations.  If we consider characterising some reference Hamiltonian ($H_r$), we can use this to define the coordinate axes and then consider a second Hamiltonian ($H_k$).  This provides a second axis to rotate around which must also be characterised and the angle $\phi$ between these two axes must be determined.  To measure this azimuthal angle, a different initialisation point must be chosen whose Bloch vector is linear independent of the original initialisation point.  A convenient choice is to rotate $\mathbf{s}$ around the first axis ($\mathbf{d}_r$) until in sits on the `equator' defined by $\theta=\pi/2$.  The second Hamiltonian is then switched on instead and the qubit precesses around $\mathbf{d}_k$.  The z-projection of this rotation can then be used to determine the angle $\phi$ between the two axes.  As $\bf{d}_r$ and $\bf{d}_k$ have already been completely characterised, the entire process can be `boot-strapped', progressively learning more information about the system.  Of course, this process of measuring different Hamiltonians is equivalent to measuring the dependence of a system Hamiltonian on the `settings of a dial' where each Hamiltonian corresponds to a different value for the input parameters.

To rotate $\mathbf{s}$  onto the equator, starting at $\bf{s}_0$ we apply $\mathbf{d}_r$ for a time 
\begin{equation}
t=\frac{1}{\omega_r}\arccos\left[\frac{\cos(2\theta_r)+1}{\cos(2\theta_r)-1}\right],
\end{equation}
which places the system in state $\mathbf{s}_1=[\cos(\beta),\sin(\beta),0]^T$ where $\beta=\arctan[-\sec(\theta_r)\sqrt{-2\cos(2\theta_r)}]$~\cite{Schirmer:04}.  If we then use this as the new initialisation point, the z-component  of the precession about $\mathbf{d}_k$ is given by
\begin{equation}\label{eq:z_phi}
z(t)=C[1-\cos(\omega_k t)]+D\sin(\omega_k t),
\end{equation}  
where $C=\frac{1}{2}\sin(2\theta_k)\cos(\phi-\beta)$ and $D=\sin(\theta_k)\sin(\phi-\beta)$.  This procedure can only be applied if $\theta_r\in[\frac{\pi}{4},\frac{3\pi}{4}]$.  If $\theta_r$ or $\theta_k$ are not within this range, a more elaborate pulsing scheme is required.  Once the two axes $\mathbf{d}_r$ and $\mathbf{d}_k$ have been characterised, measuring Eq.~(\ref{eq:z_phi}) allows both Hamiltonians to be completely reconstructed. 

Using a similar method to the previous section, the parameters $C$ and $D$ can be determined from the components of the Fourier spectrum,
\begin{eqnarray*}
\rm{DFT}^{-1}[\rm{DFT}[z_m(t)]] 	& = 	& \sum_{\nu=-N_s/2}^{N_s/2}{F(\nu) e^{i2\pi(\nu/N_s)t}} \\
		& =	& F(0)+F_R(\nu_p)e^{i2\pi\nu_pt/N_s} \\ 
		&	& +F_R(-\nu_p)e^{-i2\pi\nu_pt/N_s} \\
		&	& +iF_I(\nu_p)e^{i2\pi\nu_pt/N_s} \\
		&	& -iF_I(-\nu_p)e^{-i2\pi\nu_pt/N_s} \\
		& =	& F(0)+2F_R(\nu_p)\cos(2\pi\nu_pt/N_s)\\
		&	& -2F_I(\nu_p)\sin(2\pi\nu_p/N_s) \\
		& \simeq	& z(t).
\end{eqnarray*}
where $F_R$ and $F_I$ are the real and imaginary parts of the Fourier components.  As the measurement error of the system has already been determined from the measurements of the other axes, the constants $C$ and $D$ can be determined directly using
\begin{equation}
C=\frac{-2F_R(\nu_p)}{(1-2\eta)}\label{eq:C_phi}
\end{equation}
and
\begin{equation}
D=\frac{-2F_I(\nu_p)}{(1-2\eta)}.\label{eq:D_phi}
\end{equation}
These equations are valid if the MPP has been found exactly, though this will very rarely be the case.  Any error induced in the magnitude of the Fourier components by this effect will be small, but the error induced in the complex \emph{phase} (denoted $\chi$ so as not to be confused with the Hamiltonian angle $\theta$) will not be negligible and must be corrected.  We may do this by observing that in Eq.~(\ref{eq:z_phi}) the constant term and the negative amplitude of the cosine term must be equal.  We can define the corrected complex angle $\chi_c$ so that this is the case using
\begin{equation}
\chi_c=\arccos\left[\frac{-F(0)}{2F_R(\nu_p)}\right],
\end{equation}
such that the corrected Fourier component  
\begin{equation}
F_c(\nu_p)=|F(\nu_p)|[\cos(\chi_c)+i\sin(\chi_c)],
\end{equation}
is then used in Eqs.~(\ref{eq:C_phi}) and (\ref{eq:D_phi}).

At this point, in order to keep track of the various sine and cosine terms and their uncertainties, we will introduce the following notation.  When dealing with an angle we use $A_{\Phi}=\cos(\hat{\Phi})$ and $\delta A_{\Phi}$ to refer to the cosine of the angle and its uncertainty respectively.  Likewise, we define $B_{\Phi}=\sin(\hat{\Phi})$  as the sine of the angle giving the relationship $A_{\Phi}=\sqrt{1-B_{\Phi}^2}$ and $A_{\Phi} \delta A_{\Phi}=B_{\Phi} \delta B_{\Phi}$. 

As the value of $\theta_k$ has already been determined, $\phi$ can be found from either $C$ or $D$, depending on the value of $\theta_k$.  For instance using
\begin{equation}\label{eq:PhifromC}
\begin{array}{llll}
A_{\phi-\beta} & = & \cos(\phi-\beta) \\
	   & = & 2C/\sin(2\theta_k) \\
	   & = & C/(A_{\theta_k} B_{\theta_k}) & \theta_k>\frac{3\pi}{8},	  \\ \\
B_{\phi-\beta} & = & \rm{sgn}(D)\sqrt{1-A_{\phi-\beta}^2}  &
\end{array}
\end{equation}
or
\begin{equation}\label{eq:PhifromD}
\begin{array}{llll}
B_{\phi-\beta} & = & \sin(\phi-\beta) \\
	   & = & D/\sin(\theta_k) \\
	   & = & D/B_{\theta_k} & \theta_k<\frac{3\pi}{8}, \\ \\
A_{\phi-\beta} & = & \sqrt{1-B_{\phi-\beta}^2}, & 
\end{array}
\end{equation}
depending on the value of $\theta_k$, will minimise the effects of noise.  The angle $\phi$ is then given by 
\begin{equation}
\phi=\arccos(A_{\phi-\beta})+\beta,
\end{equation}
as expected.  

As the rotation about the axis $\bf{d}_r$ can only be performed to the same accuracy as the axis itself is characterised, there will also be some uncertainty in the angle $\hat{\beta}$.  This can be approximated by setting $\delta \theta_r \approx \delta \beta$, which gives the uncertainty
\begin{equation}
\delta A_\beta=\frac{B_\beta}{B_{\theta_r}}\delta A_{\theta_r},
\end{equation}
in $A_\beta=\cos(\hat{\beta})$.

\section{Estimating the uncertainty in $\phi$}\label{sec:phi_err}
The uncertainty in $\hat{\phi}$ will depend on the uncertainty in both the original axes characterisation and the noise in the Fourier spectrum used to compute $C$ and $D$.  The uncertainty in the parameters $C$ and $D$ can be calculated using
\begin{equation}
\delta C^2=\left|\frac{3}{2(1-2\eta)}\right|^2\delta F^2+\left|\frac{2C}{(1-2\eta)}\right|^2\delta\eta^2
\end{equation}
and
\begin{equation}
\delta D^2=\left|\frac{2}{(1-2\eta)}\right|^2\delta F^2+\left|\frac{2D}{(1-2\eta)}\right|^2\delta\eta^2
\end{equation}
where $\delta F$ and $\delta\eta$ are those defined in section~\ref{sec:theta_err}.  Here, we have ignored the covariance term to simplify the analysis.  The contribution due to correlated errors is small as the calculation of $\phi$ depends on three sets of measurements ($\bf{d}_r$, $\bf{d}_k$ and $A_{\phi-\beta}$) which are independent of each other. 

We can then define the uncertainty in $A_{\phi-\beta}$ in terms of $C$ or $D$ as
\begin{equation}
\delta A_{\phi-\beta}^2=A_{\phi-\beta}^2\left[\left(\frac{\delta C}{C}\right)^2+\left(\frac{\delta A_{\theta_k}}{A_{\theta_k}}\right)^2+\left(\frac{\delta B_{\theta_k}}{B_{\theta_k}}\right)^2\right]
\end{equation}
or
\begin{equation}
\delta A_{\phi-\beta}^2=A_{\phi-\beta}^2\left[\left(\frac{\delta D}{D}\right)^2+\left(\frac{\delta B_{\theta_k}}{B_{\theta_k}}\right)^2\right].
\end{equation}
Writing the cosine of $\hat{\phi}$ as
\begin{equation}
A_\phi=\cos(\hat{\phi})=A_\beta A_{\phi-\beta} - B_\beta B_{\phi-\beta},
\end{equation}
gives the uncertainty relationship
\begin{eqnarray}
\delta A_\phi^2 & = & \left(A_{\phi-\beta}^2+\frac{B_{\phi-\beta}^2 A_\beta^2}{B_\beta^2} \right)\delta A_\beta^2 \nonumber \\
 & & +\left(A_\beta^2+\frac{B_\beta^2 A_{\phi-\beta}^2}{B_{\phi-\beta}^2}\right)\delta A_{\phi-\beta}^2.
\end{eqnarray}

\section{Estimating the Uncertainty in the Hamiltonian parameters}\label{sec:H_err}

Once the estimates $\hat{\theta_r}$, $\hat{\theta_k}$, $\hat{\phi}$, $\hat{\omega_r}$ and $\hat{\omega_k}$ and have been found, the Hamiltonians can be estimated using the following equations,
\begin{eqnarray}\label{eq:Hrx}
\hat{H}_r & = & \frac{\hat{\omega}_r}{2}(B_{\theta_r} \sigma_x+A_{\theta_r} \sigma_z) \nonumber \\
		& = & H_{r,x}\sigma_x+H_{r,z}\sigma_z
\end{eqnarray}
and
\begin{eqnarray}
\hat{H}_k & = & \frac{\hat{\omega}_k}{2}(B_{\theta_k} A_\phi \sigma_x + B_{\theta_k} B_\phi \sigma_y + A_{\theta_k} \sigma_z) \nonumber \\ 
		& = & H_{k,x}\sigma_x+H_{k,y}\sigma_y+H_{k,z}\sigma_z,
\end{eqnarray}
where $H_{j,i}$ is the $i$-th component of the $j$-th Hamiltonian.
The uncertainty in each of the components of $H_r$ are given by
\begin{equation}
\left(\frac{\delta H_{r,x}}{\hat{H}_{r,x}}\right)^2=\left(\frac{\delta B_{\theta_r}}{B_{\theta_r}}\right)^2+\left(\frac{\delta \omega_r}{\hat{\omega}_r}\right)^2
\end{equation}
and
\begin{equation}
\left(\frac{\delta H_{r,z}}{\hat{H}_{r,z}}\right)^2=\left(\frac{\delta A_{\theta_r}}{A_{\theta_r}}\right)^2+\left(\frac{\delta \omega_r}{\hat{\omega}_r}\right)^2.
\end{equation}
For $H_k$ the component uncertainties are 
\begin{equation}
\left(\frac{\delta H_{k,x}}{\hat{H}_{k,x}}\right)^2=\left(\frac{\delta B_{\theta_k}}{B_{\theta_k}}\right)^2+\left(\frac{\delta \omega_k}{\hat{\omega}_k}\right)^2+\left(\frac{\delta A_\phi}{A_\phi}\right)^2,
\end{equation}
\begin{equation}
\left(\frac{\delta H_{k,y}}{\hat{H}_{k,y}}\right)^2=\left(\frac{\delta B_{\theta_k}}{B_{\theta_k}}\right)^2+\left(\frac{\delta \omega_k}{\hat{\omega}_k}\right)^2+\left(\frac{\delta B_\phi}{B_\phi}\right)^2
\end{equation}
and
\begin{equation}\label{eq:Hkz}
\left(\frac{\delta H_{k,z}}{\hat{H}_{k,z}}\right)^2=\left(\frac{\delta A_{\theta_k}}{A_{\theta_k}}\right)^2+\left(\frac{\delta \omega_k}{\hat{\omega}_k}\right)^2.
\end{equation}

\section{Example Simulations}\label{sec:ExampleSims}
\begin{figure} [tb!]
\centering{\includegraphics[width=8cm]{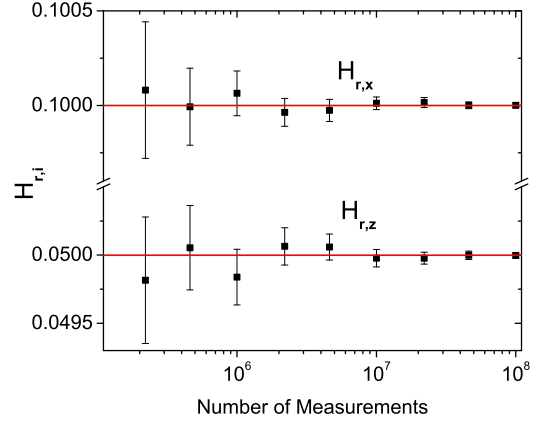}} \caption
{An example of the systematic reduction in the uncertainty of the Hamiltonian parameters as the number of measurements is increased.  The error bars are given by three times the uncertainty estimate for each point and the solid line gives the `true' value ($H_{r,x}=0.1$, $H_{r,z}=0.05$).  The estimates are seen to converge to the true value as the number of measurements are increased.\label{fig:Hr_example}}
\end{figure}

To illustrate these ideas and determine the accuracy of the parameter estimate and its uncertainty, we simulated the measurement procedure on an arbitrary example system, $H_r=0.1\sigma_x+0.05\sigma_z$.  Using an observation time $t_{\rm{ob}}=500$ and progressively larger numbers of measurements, the increase in precision can be observed.   In Fig.~\ref{fig:Hr_example}, the components $H_{r,x}$ and $H_{r,z}$ are plotted for increasing numbers of measurements.  The errors bars are given by $3\delta H$ which should be equivalent to the 3-sigma level and the true value is shown as a solid line.  As the number of measurements increases, the uncertainty reduces and the estimated values converge to the true value, as expected.  The complete process is then simulated using a second example Hamiltonian ($H_k=0.6\sigma_x+0.45\sigma_y+0.1\sigma_z$) and similar results are obtained but with increased uncertainty as the components of $H_k$ rely on the measurements of both $H_r$ and $H_k$, so there is more scope for accumulated errors.

In order to compare the uncertainty calculated using the equations in section~\ref{sec:H_err} with the expected spread of the data, we repeated the simulations of the example system many times with the same number of measurements.  By looking at the spread of the resulting estimates from many experiments and comparing this to the derived uncertainty from one experiment we can confirm that the uncertainty provides a good bound.  Providing a good error bound on the Hamiltonian parameters alleviates the need to perform characterisation many times to obtain good statistics.

\subsection{Accuracy of the Uncertainty Estimate}\label{sec:AccuracyTests}
To measure the distance between the real Hamiltonian vector $\bf{d}$ and its estimate $\hat{\bf{d}}$ we use the following distance metric,
\begin{equation}\label{eq:HamilDist}
\mathcal{D}=\frac{|\bf{d}-\hat{\bf{d}}|}{|\bf{d}|}
\end{equation}
and a measure of the uncertainties is 
\begin{equation}\label{eq:HamilUncertDist}
\delta\mathcal{D}=\frac{\sqrt{\delta d_x^2+\delta d_y^2+\delta d_z^2}}{|\hat{\bf{d}}|}=\frac{|\delta \bf{d}|}{|\hat{\bf{d}}|}.
\end{equation}

\begin{figure} [tb!]
\centering{\includegraphics[width=7.5cm]{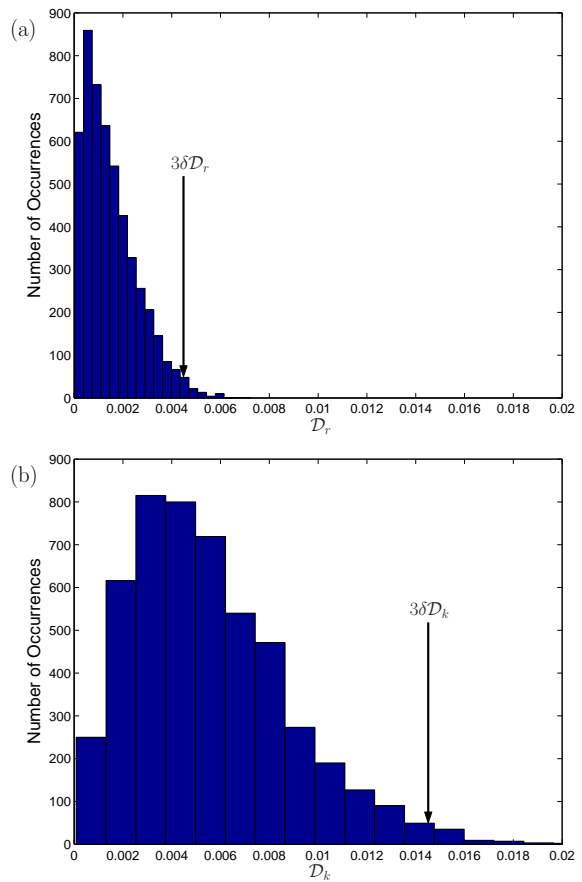}} \caption
{The distribution of $\mathcal{D}$ for the estimated (a) $H_r$ and (b) $H_k$ over 5000 simulated runs.  For these simulations, (a) $98.4\%$ and (b) $98.7\%$ of the estimates are found to lie within the average uncertainty interval ($3\overline{\delta \mathcal{D}}$).  The absolute uncertainty in $H_k$ is greater than for $H_r$ as more steps are required, giving a larger accumulated error.\label{fig:Fr_combined}}
\end{figure}

We simulated the characterisation procedure for the example system using $t_{\rm{ob}}=500$, $N_{\rm{s}}=10000$ and $N_e=50$ with a measurement error probability of $10\%$ ($\eta=0.1$).  Fig.~\ref{fig:Fr_combined}(a) shows a histogram of $\mathcal{D}$ for $H_r$ over 5000 simulated runs, the average uncertainty $\overline{\delta\mathcal{D}}$ over 5000 runs is also shown.  For this example $98.4\%$ of the simulation runs lie within $3\overline{\delta\mathcal{D}}$, illustrating that the uncertainty provides a good bound on the estimated parameters.  

The fidelity $\mathcal{D}$ for $H_k$ shows a similar distribution, though the absolute uncertainty is greater for a given number of measurements as more steps are required to determine the azimuthal angle $\phi$.  Fig.~\ref{fig:Fr_combined}(b) shows the equivalent histogram for determination of the Hamiltonian $H_k$ over 5000 simulated runs.  Three times the average uncertainty ($3\overline{\delta \mathcal{D}}$) includes $98.7\%$ of the data.  The intervals for both $\mathcal{D}_r$ and $\mathcal{D}_k$ are slightly too small as a 3-sigma interval should contain approximate $99.7\%$ of the data.  This discrepancy is due to the effect of correlated errors between the Fourier components $\delta F$ and the uncertainty in the MPP location ($\delta \omega$).  In general, as the noise level in the Fourier spectrum increases, the width of the peak $P(t_p)$ will also increase.  This results in a small correlation between the uncertainties in $\hat{\omega}$ and $\hat{\theta}$ which has not been taken into account.  For a given set of experimental data, the width of $P(t_p)$ and the standard deviation of the the noise floor of the Fourier spectrum will decrease as the number of measurements increases.  The relationship between these errors can then be determine and will be (in general) non-trivial.  The covariance can then be calculated, the result of which would be to add an additional term to Eqs.~(\ref{eq:Hrx})-(\ref{eq:Hkz}) and therefore increasing the overall uncertainty.  This additional term will be small as the fractional uncertainty in $\hat{\omega}$ is typically much smaller than in $\hat{\theta}$ which implies the covariance between them will also be small, relative to the other uncertainties. 

\begin{figure} [tb!]
\centering{\includegraphics[width=7cm]{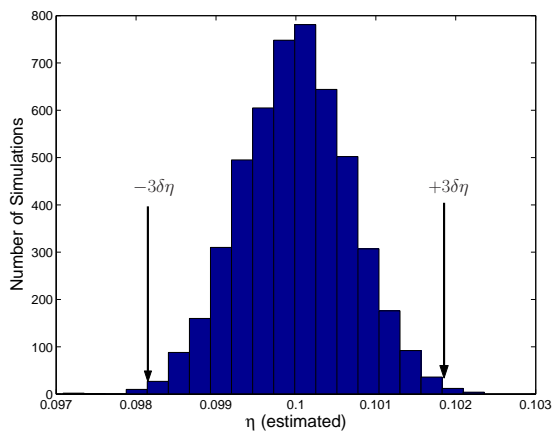}} \caption
{The distribution of the estimated measurement error ($\hat{\eta}$) for 5000 simulated runs.  For this simulation, $99.5\%$ of the estimates lie within the uncertainty $\pm3\delta \eta$.\label{fig:eta_est}}
\end{figure}

The measurement error estimate ($\hat{\eta}$) is found to be very well behaved, with $99.5\%$ of the estimates lying within the error bounds, which is very close to what is expected for a 3-sigma confidence interval.  A histogram of $\hat{\eta}$ is shown in Fig.~\ref{fig:eta_est} with $3\overline{\delta\eta}$ labelled for 5000 runs, each run consisting of $N_T=5\times10^5$ measurements.

\subsection{Scaling behaviour of the Uncertainty}\label{sec:Scaling}
\begin{figure} [tb!]
\centering{\includegraphics[width=6.5cm]{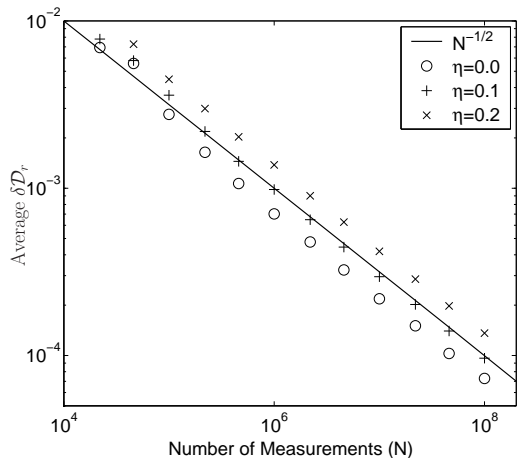}} \caption
{The average uncertainty $\overline{\delta\mathcal{D}_r}$ of the estimate for the Hamiltonian $H_r$ as a function of total number of measurements.  Each data point is the average of 10 simulation runs.  The solid line shows $1/\sqrt{N}$ where $N$ is the number of measurements.  As the total number of measurements increases, the overall precision with which the Hamiltonian is know increases.  For a random measurement error, the achievable precision is reduced but still asymptotes to a scaling of one over the square-root of the number of measurements.\label{fig:Fr_scaling}}
\end{figure}

The usefulness of this technique is ultimately governed by how many measurements are required to obtain a given precision in the final Hamiltonian estimate.  To investigate this, the example system was characterised with progressively larger numbers of measurements.  The average of the resulting estimated uncertainty $\overline{\delta\mathcal{D}_r}$ is plotted in Fig.~\ref{fig:Fr_scaling} for several different values of the measurement error ($\eta$).  For increasing numbers of measurements, the Hamiltonian estimate gets progressively more accurate, as expected.  This scaling is approximately proportional to $1/\sqrt{N}$ with the achievable precision reduced by the effect of the measurement error.  This constant factor is effectively a `penalty' which depends on the measurement error but is largely independent of the number of measurements.  

From this type of analysis we can estimate how many measurements are required to achieve a certain precision in the final result.  Assuming all other factors are negligible, the achievable precision scales as one over the square-root of the number of measurements.  Other factors, such as control field fluctuations, will ultimately limit this process.  This is easily identified as the achievable precision will tend to asymptote to some value which is limited by these fluctuations.

\section{Implications for Single-qubit rotations in Quantum Computing}\label{sec:compositepulses}
In order to be able to perform single-qubit rotations of the type required for quantum computing applications, a certain level of accuracy is required.  The threshold theorem for quantum error correction states that if a physical error rate of $p=10^{-4}-10^{-5}$ can be achieved then concatenated quantum error correction protocols can be implemented successfully for arbitrary precision computation\cite{Preskill:98}.  This physical error rate gives the probability of a discrete error due to decoherence of the system.
The errors introduced due to inaccurate characterisation will also contribute, though in a less predictable way.  Typically, gate operations are assumed to have a precision of $10^{-6}$ or better but from the previous analysis, this would require $10^{12}$ measurements during characterisation.  For a typical measurement readout time of $1\rm{\mu s}$ this gives an initial characterisation time of approximately 12 \emph{days}.

This turns out to be an overly simplistic view as the precision of the gate operations is not equivalent to the probability of a discrete error due to decoherence.  For a single gate rotation around an ideal angle $\theta$, the true rotation will be around an angle $\theta(1+\epsilon)$ and therefore the probability of a discrete error $p\propto (\epsilon)^2$ where $\epsilon \approx \delta\theta/\theta$.  Given the previous discussion on the scaling of $\delta\theta$ with number of characterisation measurements $N$, the probability of discrete error on a single gate operation actually scales proportional to $N^{-1}$, which requires only $10^6$ rather than $10^{12}$ measurements.  

As well as errors induced by inaccurate knowledge of the Hamiltonian angle ($\theta$), errors can also be introduced due to an inaccurate rotation frequency or `over rotation error'.  In general this will have a similar effect to an angle characterisation error as (for small errors) they are equivalent.  In addition, for the characterisation process discussed in this paper, the percentage uncertainty in the rotation frequency is typically an order of magnitude smaller than the uncertainty in the Hamiltonian angle which means that angle errors are the dominant source of gate error.

For multiple gate operations, the probability of a discrete error scales as $np$ where $n$ is the number of gate operation time steps and therefore the number of possible error locations\cite{Knill:98}, assuming that errors in different qubits are uncorrelated.  In the worst case, the rotation error accumulates as $n\epsilon$ which gives $p_T=np\propto (n\epsilon)^2$, the total probability of error for $n$ possible error locations.  This means its possible (in the worst case) for the uncertainty in the angle to accumulate over multiple rotations.  This will not always be the case as certain rotations (such as a $2\pi$ rotations) are less susceptible to characterisation errors than others and it is possible to get error cancellation.  While this discussion is not new\cite{Preskill:98,Knill:98}, the $1/\sqrt{N}$ scaling of the achievable precision in $\delta\theta$ highlights the very real constraints imposed by the measurement and therefore characterisation time of any prospective quantum computing proposal.

Several techniques exist for dealing with characterisation errors of this kind~\cite{Vandersypen:04,Wu:03}, much of which has recently regained interest for QIP applications \cite{Jones:03,Cummins:03}.  One such technique, which has been known in the NMR literature for some time, is composite pulsing\cite{Ernst:90}.  This involves carefully constructing a pulse sequence for a given rotation in order to reduce characterisation errors in both the angle (off-resonant errors) and the rotation frequency (pulse length errors).  Recent work by Brown et al.\ \cite{Brown:04} has shown that in-fact, systematic characterisation errors can be eliminated to arbitrary order using strings of composite pules.  For a single imperfect gate with fractional error $\epsilon$, the resulting gate error can be reduced to $\mathcal{O}(\epsilon^n)$ for arbitrary $n$ using a composite pulse sequence whose length scales as $n^3$.  Using this or similar techniques, we can imagine a trade-off between long initial characterisation time (large number of characterisation measurements) and longer composite pulse sequences for our gate operations (slower operating speed).  In addition, by choosing fine time sampling (large $N_s$) we can obtain very precise frequency estimates at the expense of poor angular resolution due to small numbers of ensemble measurements ($N_e$).  The imprecise angular estimate could then be accounted for using composite pulsing.  Similarly, poor time resolution and large numbers of ensemble measurements will give accurate angle estimates at the expense of rotation frequency resolution. There may also be situations where it is advantageous to precisely characterise some gates and/or qubits but not others. 

\section{Conclusion}
As the precision and level of complexity of quantum control experiments increases, the accuracy to which pertinent system parameters are known must also increase.  While this is most commonly discussed in the context of quantum computing, the ability to precisely measure the terms in an arbitrary Hamiltonian has much broader application to the study of quantum systems. 

The procedure given here for characterising an arbitrary two-state Hamiltonian has distinct advantages over other methods.  Given only one measurement axis and assuming the system can be repeatedly initialised in a single state which is not an eigenstate of the Hamiltonian to be characterised, the Hamiltonian parameters can be determined to arbitrary accuracy.  By taking the discrete Fourier transform of a series of measurements of the evolution of the system, the parameters in the Hamiltonian can be computed directly from the Fourier components.  

Using signal processing techniques, the uncertainty in the Hamiltonian parameters can be estimated and we have derived example expressions for these uncertainties.  If a random measurement error is present, this too can be characterised with an uncertainty.  This uncertainty estimate is found to scale proportionally to one on the square-root of the total number of measurements.  The introduction of measurement error reduces the achievable precision by a constant factor which is independent of the number of measurements.

In the laboratory, this procedure can be applied as the experiment progresses, giving an increasing more accurate estimate of the parameters in question.  It also means that if the response of a Hamiltonian to a given input parameter is required, as the input parameter is varied, the resulting system can be determine with an uncertainty at each point.  This enables the usual (non-)linear fitting routines to be applied to the problem to find the general response function.

Being able to accurately characterise a Hamiltonian is vitally important if we are to move beyond proof-of-concept experiments and build working devices for QIP.  The trade-off between more accurate initial characterisation and more sophisticated gate sequences allows these devices to be optimised for a particular application.
 
\begin{acknowledgments}
JHC would like to acknowledge helpful discussions with S.~J.~Devitt.  This work was supported in part by the Australian Research Council, the Australian government, the US National Security Agency, the Advanced Research and Development Activity and the US Army Research Office under contract number DAAD19-01-1-0653.  SGS and DKLO acknowledge funding from the Cambridge-MIT institute, Fujitsu, EPSRC and EU grants TOPQIP and RESQ.  DKLO also thanks Sidney Sussex College for support.
\end{acknowledgments}

\appendix
\section{Derivation of the time evolution of $\langle\sigma_z\rangle$ under an arbitrary two-state Hamiltonian}\label{app:sigz_derv}
Given an arbitrary two-state Hamiltonian, we can write it in terms of the Pauli matrices using Eq.~(\ref{eq:Handd}).  The free evolution of the system under this Hamiltonian is given by the operator $U(t)=e^{-iHt}$ which, using a generalised de Moivre formula~\cite{Merzbacher:98}, can be rewritten as
\begin{equation}
U(t)=e^{-id_ot/2}\left[\mathbf{I}\cos\left(\frac{|\bf{d}|t}{2}\right)-i\hat{\mathbf{d}}.\vec{\sigma}\sin\left(\frac{|\bf{d}|t}{2}\right)\right].
\end{equation}
If the system is initially in the state $|\psi(0)\rangle=|0\rangle$ ($\theta=\phi=0$) then (converting to polar coordinates) the evolution of the system is given by
\begin{widetext}
\begin{equation}
|\psi(t)\rangle=U(t)|\psi(0)\rangle=e^{id_0t/2}\left\{ \left[\cos\left(\frac{|\bf{d}|t}{2}\right)-i\cos\theta\sin\left(\frac{|\bf{d}|t}{2}\right)\right]|0\rangle+\sin\theta\sin\left(\frac{|\bf{d}|t}{2}\right)(\sin\phi-i\cos\phi)|1\rangle\right\}.
\end{equation}
\end{widetext}
The observable in this case is the projection onto the z-axis so we will use $\acute{z}=|0\rangle\langle0|-|1\rangle\langle1|$ as the operator which gives the expectation value of the z-projection,
\begin{equation}
\langle\sigma_z\rangle=\langle\acute{z}\rangle=\langle\psi(t)|\acute{z}|\psi(t)\rangle.
\end{equation}	
After cancelling the global phase and rearranging terms, this becomes
\begin{equation}
\langle\sigma_z\rangle=\cos^2\left(\frac{|\bf{d}|t}{2}\right)+(\cos^2\theta-\sin^2\theta)\sin^2\left(\frac{|\bf{d}|t}{2}\right).
\end{equation}
If we set $|\bf{d}|=\omega$, the angular frequency of the precession, this gives the time dependence of the z-projection
\begin{equation}
z(t)=\langle\sigma_z\rangle=\cos\omega t\sin^2\theta+\cos^2\theta,
\end{equation}
as expected.

\newpage 
\bibliography{hc1q}

\end{document}